\documentclass[a4paper]{article}

\usepackage{18lomcon}        % Proceedings volume layout metrics
\usepackage{cite}             % Smart range citations
\usepackage{epsfig}           % Encapsulated PostScript figure inclusion
\usepackage{epstopdf}

\usepackage{amssymb}
\usepackage{amsmath} 
\usepackage{amssymb} 
\usepackage{xspace}

\usepackage[left]{lineno}
\newcommand{\nubare}{\ensuremath{\overline{\nu}_e}\xspace}
\newcommand{\nubarenubare}{\ensuremath{\nubare \rightarrow \nubare\xspace}}

\begin{document}

%%%%%%%%%%%%%%%%%%%%%%%%%%%%%%%%%%%%%%%%%%%%%%%%%%%%%%%%%%%%%%%%%%
% The preamble of the paper
%%%%%%%%%%%%%%%%%%%%%%%%%%%%%%%%%%%%%%%%%%%%%%%%%%%%%%%%%%%%%%%%%%

\title{STATUS AND PHYSICS POTENTIAL OF THE JUNO EXPERIMENT}

\author{Giuseppe Salamanna \email{salaman@fis.uniroma3.it}, {\it on behalf of the JUNO Collaboration}}

\affiliation{Roma Tre University and INFN Roma Tre, 00146 Rome, Italy}

% You may repeat \author and \affiliation as many times as necessary!

\date{}
% Print it out!
\maketitle

%%%%%%%%%%%%%%%%%%%%%%%%%%%%%%%%%%%%%%%%%%%%%%%%%%%%%%%%%%%%%%%%%%
% The preamble of the paper
%%%%%%%%%%%%%%%%%%%%%%%%%%%%%%%%%%%%%%%%%%%%%%%%%%%%%%%%%%%%%%%%%%

\begin{abstract}
The Jiangmen Underground Neutrino Observatory (JUNO) is
an underground 20 kton liquid scintillator detector being built
in the south of China and expected to start data taking in
2020. JUNO has a physics programme focused on neutrino
properties using electron anti-neutrinos emitted from two
near-by nuclear power plants. Its primary
aim is to determine the neutrino mass hierarchy from the $\nubare$
oscillation pattern.
With an unprecedented relative energy resolution of 3$\%$ as
target, JUNO will be able to do so
with a statistical significance of 3-4 $\sigma$ within six years of
running. It will also measure other
oscillation parameters to an accuracy better than 1$\%$.
An ambitious experimental
programme is in place to develop and optimize the detector and
the calibration system, to maximize the light yield and
minimize energy biases. JUNO will also be in a good position to study neutrinos
from the sun and the earth and from supernova explosions, as
well as provide a large acceptance for the search for proton
decay.
JUNO's physics potential was described and the status of
its construction reviewed in my talk at the conference.
\end{abstract}

\section{Introduction}

The Jiangmen Underground Neutrino Observatory (JUNO) is a neutrino 
experiment being built in China, described in \cite{An:2015jdp}. 
Its primary purpose is to determine the neutrino mass hierarchy (MH) 
and measure the oscillation parameters using reactor sources. It will
consist of a large mass, pure liquid scintillator (LS), placed in an acrylic 
sphere of 35.4 m of diameter; a system of 
large-area photo-multipliers of new generation (PMT); a veto system.
It will be located at a distance of approximately 50 km from two power plants
 (Yangjiang and Taishan). The two plants are expected to provide
an equal thermal power of about 18 GW, but at the start-up of 
the experiment only 26.6 GW are expected to be available. 
The baseline is optimized for maximum $\nubare$ disappearance, 
i.e. a minimum of the survival probability (here in the case of 
normal neutrino mass hierarchy (NH)): 
\begin{eqnarray}\label{eq:1}
P_{NH}&(\nubarenubare)=&1-\frac{1}{2}\sin^2 2\theta_{13}\left(1-\cos\frac{L\Delta m^2_{atm}}{2E_{\nu}}\right)  \\
&& -\frac{1}{2}\cos^4 \theta_{13}\sin^2 2\theta_{12}\left(1-\cos\frac{L\delta m^2_{sol}}{2E_{\nu}}\right) \nonumber \\
+\frac{1}{2}&\sin^2 2\theta_{13}\sin^2 \theta_{12}&\left(\cos\frac{L}{2E_{\nu}}\left(\Delta m^2_{atm}-\delta m^2_{sol}\right)
-\cos\frac{L\delta m^2_{atm}}{2E_{\nu}}\right), \nonumber
\end{eqnarray}
(the probability for the inverted hierarchy $P_{IH}$ differing only 
in the coefficient of the last element in the sum,
$\cos^2\theta_{12}$). JUNO will be placed about 700 m below underground, correspondinge to 
about ~1900 m.w.e., in a pit dug in the ground afresh. 
To pursue its main physics goals JUNO will need to attain 
an unprecedented resolution 
on the energy of the $\nubare$ produced in the reactors. 
In order to meet such performance extensive studies have been conducted
over the past few years concerning various aspects of engineering, 
detector design, including 
optical/light collection in the LS and PMT and response of the read-out electronics, 
software development and background suppression \cite{junoCDR}. In the talk given at the 
18th Lomonosov Conference on Elementary Particle Physics, the main physics case 
of JUNO was described; the status of the detector design optimization and 
construction was illustrated and some examples of physics measurements possible, 
beyond the neutrino oscillations, were presented. 

\section{What drives the detector design....}
\begin{figure}[t]
     \centering
     \begin{center}
       \mbox{
         \epsfig{file=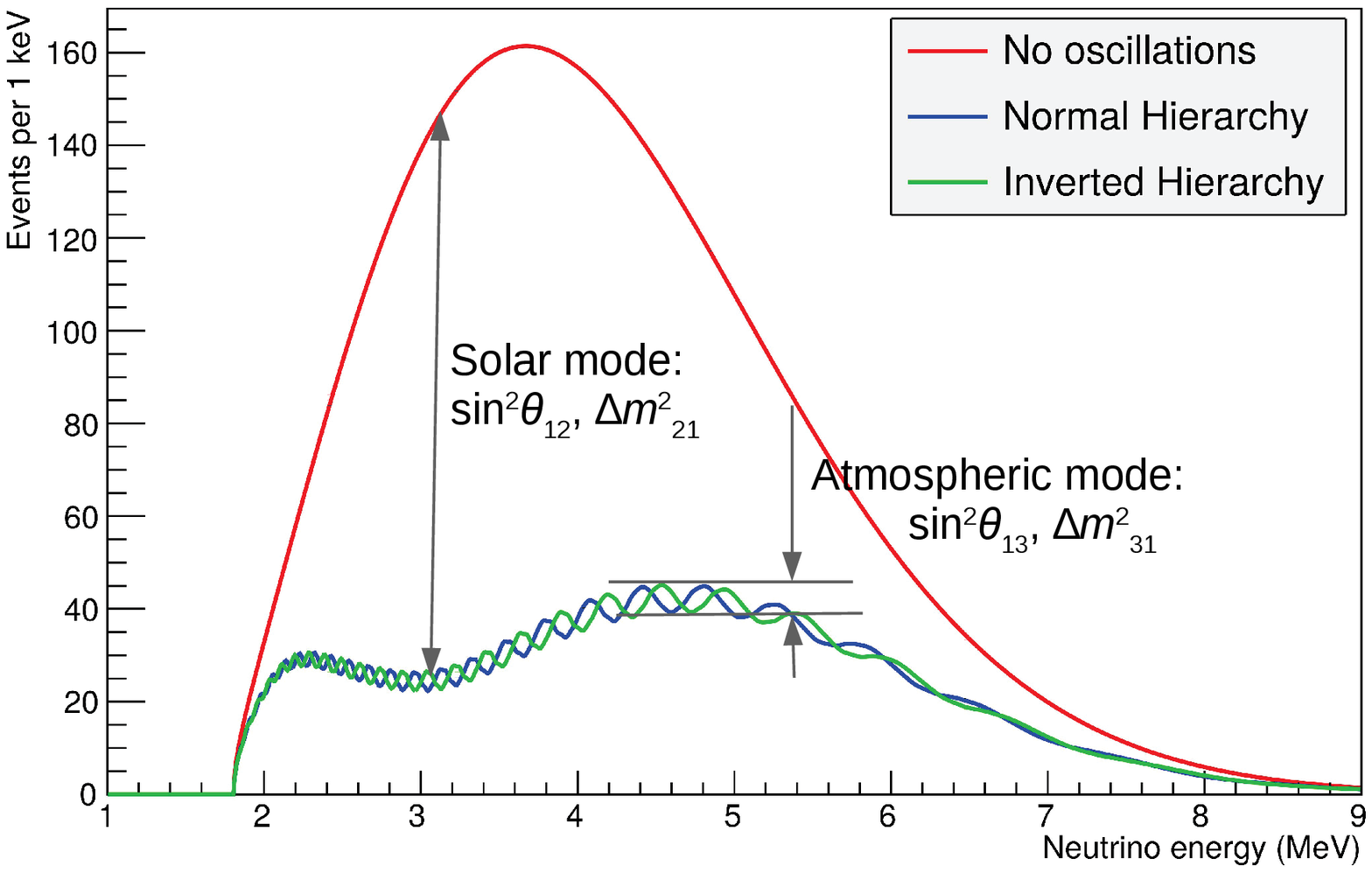,height=5 cm}
         \epsfig{file=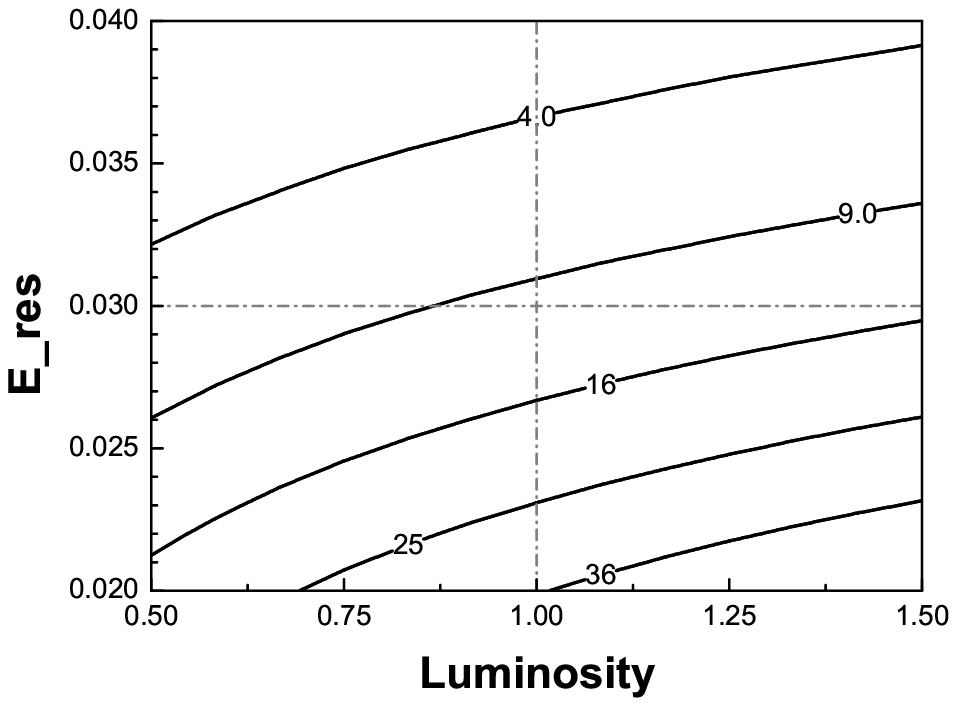,height=5.2 cm}
       }
     \end{center}
     \caption{Left: $P_{MH}(\nubarenubare)$ for no oscillation and oscillations under the two MH hypothesis (courtesy: Y.Malyshkin), just for illustration purposes. Right: curves of $\Delta\chi_{min}^2 (NH-IH)$ as a function of the energy resolution (y axis) and the ``luminosity'' of the sample collected with respect to the 6-year benchmark \cite{An:2015jdp}}
     \label{fig:1}
\end{figure}

The experimental signature of the reactor $\nubare$ in the JUNO detector
 is given by the inverse beta decay (IBD) process 
$\ensuremath{\overline{\nu}_e}\xspace + p \rightarrow e^+ + n$,
where the $p$ and $n$ are a proton from the LS and a neutron, respectively.
The resulting signal is given by a visible energy from the positron energy loss and 
annihilation, plus delayed light at a fixed 2.2 MeV energy from the neutron capture.
The main goal of JUNO is to determine the MH with at least a 3 $\sigma$ significance 
within the first 6 years of data taking. The correct MH will be extracted by means 
of a $\chi^2$ fit to the kinetic energy spectrum
of the prompt $e^+$ ($T_{e^+}$), which is directly related to the 
$\nubare$ energy ($E_{\nubare} = T_{e^+} + 2\times m_e + 0.8$ MeV, where $m_e$ is the 
$e^+$ mass). 
From Eqn. \ref{eq:1} one sees that the two hierarchies have a difference in the fine structure of
$P_{MH}(\nubarenubare)$, which is illustrated in Fig. \ref{fig:1}, left, and was pointed out in \cite{petcov}. 
The correct MH is determined by constructing the $\Delta\chi_{min}^2 (NH-IH)$ from 
the two fits to the experimental reactor data; this can be translated into a statistical 
significance of the discrimination.
It should be noticed that, with such strategy, a residual ambiguity lingers, associated to the correct 
value of the atmospheric mass difference $\Delta m^2_{atm}$; this reduces the 
final sensitivity of the fitting procedure. 
JUNO estimates that, in order to reach the desired significance, the most important fogure of
performance is the overall resolution on the event-by-event measurement of $E_{\nubare}$.
In Fig. \ref{fig:1}, right, 
this relationship appears clearly. It is therefore crucial to design a detector which minimizes
the statistical uncertainty from stochastic fluctuations in the scintillation light collection, 
yet keeping linearity and uniformity of the energy response under control.
A 3$\%$ overall relative energy resolution will yield, for 6 years of data at 36 GW of 
reactor thermal power, a median significance of 3.4 (3.5) $\sigma$ for NH (IH) \cite{An:2015jdp}. 

\section{...and the resulting detector design}
To attain this level of precision, the JUNO collaboration has developed a detector made of 
3 basic parts, plus the electronics. The central detector is a 20 kton LS target mass, 
conceived to maximize the photon statistics and minimize the attenuation 
of the IBD prompt signal. This will be the largest volume of LS to date, composed of a mixture 
of $>98\%$ LAB (solvent, ~1200 photo-electrons/MeV), PPO (solute) and a less-than-per-cent 
part of bis-MSB (wavelength shifter). The photons are collected by PMT of two different kinds: 
about 18000 20 inch PMT, most of which of the micro-channel plate 
 (MCP-PMT) type, will guarantee an extended photo-coverage ($75\%$, as per JUNO requirement)
and a high overall detection efficiency (expected by the JUNO specifications to be $27\%$ at $\lambda$=420 nm); 
their transit time spread (TTS) being 12 ns. 25000 
“conventional” 3 inch will allow to follow a multi-calorimetric approach, whereby the 
non-stochastic terms in the energy resolution will 
be monitored during the calibration runs with known radio-active sources at different 
energies; the 3 inch PMT will extend the dynamical range in the waveform of large numbers 
of photo-electrons hitting a localized region and improve time and vertex resolution 
for muon reconstruction (against cosmic muons). 
Finally, a veto system will be in place to screen off incoming muons and photons by 
means of a surrounding water buffer (Cherenkov) and top scintillators. The project of the front-end 
electronics is also a challenge, because of the many read-out channels, the dark noise rate of the 
20 inch PMT and the needed resiliance and efficient heat dissipation of an under-water system. A large effort from the collaboration is being devoted to this task, but it will not be described here.

Such design, especially the large mass and control of the energy scale, will also allow 
JUNO to perform measurements of the neutrino solar parameters with uncertainties well below 1$\%$, not attained to date. 

\section{Current status of the detector project}
\begin{figure}[t]
     \centering
     \begin{center}
       \mbox{
         \epsfig{file=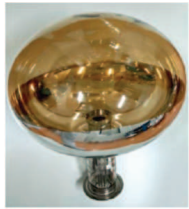,height=5 cm}
         \epsfig{file=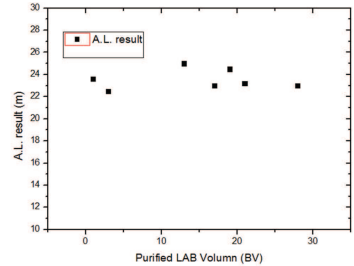,height=5cm}
       }
     \end{center}
     \caption{Left: the picture of a 20 inch MCP-PMT (see text). Right: Attenuation length vs level of LAB purification from the Daya Bay purification pilot plant.} 
     \label{fig:2}
\end{figure}

After a careful design phase, the construction of the various elements is underway. 
About 15000 MCP-PMT were ordered from the NNVT (North Night Vision of Technology CO., LTD, China) \cite{bib:nnvt}
manufacturer and are being tested at 
a dedicated centre in the region around the JUNO site. Further technical 
details on the design and production of the MCP-PMT was the subject of a dedicated 
contribution by S. Qian at this conference. Additional 5000 “conventional dynode” 20 inch 
PMT were commissioned to Hamamatsu \cite{bib:hama}, of the type R12860, in order to complement the 
leading PMT lay-out and provide a faster TTS (3 ns). All the large-area PMT will be 
equipped with protective masks to reduce propagation of shock waves if one PMT 
explodes under water pressure: their design has been finalized after extensive 
pressure tests. The bidding of the 3 inch PMT was completed last spring and
 the elements ordered from HZC-Photonics. These are custom-made based on the KM3NeT
design, with improved TTS for better muon tracking. It is desirable that the LS be purified 
to a good degree from radioactive isotopes, to reduce the intrinsic background of
the detector. The set level are $10^{-15}$ g/g for $^{238}$U and $^{232}$Th and 
$10^{-17}$ g/g for $^{40}$K. The attenuation length (AL) that JUNO requires is 
greater than 20 m at $\lambda=$ 430 nm (for 3 g/l of PPO in LAB). 
A strategy has been developed by JUNO aimed at obtaining an optimal 
admixture of solvent and
solutes in optical and radio-active terms. The purification will go through four parallel 
and complementary processes: an Al$_2$O$_3$ (alumina) column, a distillation plant, 
water extraction and gas stripping. 
A pilot plant has been established in one of the LS halls of the Daya Bay experiment
in China to monitor the AL and level of purification, which uses the alumina method. 
Results are displayed in Fig. \ref{fig:2} and show a good stability and that the desired 
level of AL has been surpassed. 

\section{Calibration of energy and vertex position reconstruction}
As stressed, a good and understood energy response is paramount for the main aims of the JUNO experimental programme. 
While the large mass, photo-coverage and detection efficiency ensure small statistical fluctuations 
in the light yield, systematic misestimation from non linearity and non uniformity of the
detector response can quickly spoil the precision.  Keeping the uncertainty on the energy scale 
determination at less than 1$\%$ over the energy interval is crucial to keep the total 
$\sigma(E)/E \leq$ 3$\%$. 
Together with that, reconstructing the position at which the event took place in the LS will
 play a very useful role in
suppressing the backgrounds, mainly related to diffuse radioactivity and cosmic ray muon scattering. 
To this end JUNO will deploy a redundant calibration
system. Complementary methods
are envisaged across the detector and for various energy loss mechanisms: an Automatic Calibration Unit (ACU)
will place several known radioactive sources along the vertical axis in the sphere; a Cable Loop System 
(CLS) will move 
across vertical planes by means of pulleys, while a Guide Tube Calibration System (GTCS) will be in use
to probe the outer CD surface. Finally a Remotely Operated under-LS Vehicle (ROV) will provide 
events of known energy across the whole detector volume. The calibration strategy (in particular
number of points in the scan and occurrence of the calibration scans) is being worked out. 

\section{Other physics possible with JUNO}
JUNO's features make it an excellent detector also for other physics topics, 
including non-reactor neutrino measurements such as
solar and supernova neutrino fluxes and geo-neutrino isotopical origin. 
In many of these case, the main challenge will be to control the intrinsic background activity and 
the cosmogenic cascades (muons interacting with carbon in the LS and creating Lithium). The latter are 
expected to be of high occurrence in the relatively shallow JUNO pit (about 250000 per day as opposed to
60 IBD events, tens to thousands of solar neutrinos and about 1 geo-neutrino per day). 
 JUNO will also be complementary to large Cherenkov detectors (e.g. SK, HK) in the search for proton decays. 
The proton will come from the hydrogen in the LS and should decay to a neutrino and a kaon, with subsequent 
semi-leptonic meson decays. The initial decay is sub-threshold for Cherenkov light in water but the kaon kinetic
energy of about 105 MeV is well visible as scintillation light. Thanks to the peculiar time pattern of the tiered 
decay, JUNO will be competitive (and complementary) soon after switching on.
Additional information on all the programme can be found in \cite{An:2015jdp}.

\section{Conclusions}
The JUNO experiment is on course to start operations within the next few years. Its challenging and multi-faceted
physics programme (on and beyond reactor neutrino oscillations) demands a very careful detector design and the use 
of novel technologies. At the same time, with its unprecedented size and energy resolution, JUNO is poised to
have an impact on many areas of neutrino physics.
A few selected figures were provided drawn from the many design/technical improvements and tests 
being put in place to achieve both unprecedented performance and reliability for a large and underwater system.

\section*{Acknowledgments}

The speaker would like to warmly acknowledge the organizers of the conference for their kindness and the high standard of the physics programme. Also, a heartly thank you for giving me the chance to spend a wonderful time in Moscow and practice the language. 

%%%%%%%%%%%%%%%%%%%%%%%%%%%%%%%%%%%%%%%%%%%%%%%%%%%%%%%%%%%%%%%%%%
% References
%%%%%%%%%%%%%%%%%%%%%%%%%%%%%%%%%%%%%%%%%%%%%%%%%%%%%%%%%%%%%%%%%%


\begin{thebibliography}{99}

\bibitem{An:2015jdp}
  F.~An {\it et al.} [JUNO Collaboration],
  %``Neutrino Physics with JUNO''
  J.\ Phys.\ G {\bf 43}, 030401 (2016)
%  doi:10.1088/0954-3899/43/3/030401
%  [arXiv:1507.05613].

\bibitem{junoCDR} 
  T.~Adam {\it et al.} [JUNO Collaboration],
  %``Neutrino Physics with JUNO''
  arXiv:1508.07166

\bibitem{petcov} S. T. Petcov and M. Piai, Phys. Lett. B {\bf 533} 94 (2002) [hep-ph/0112074].

\bibitem{bib:nnvt} North Night Vision of Technology, China: 
http://en.nvt.com.cn/

\bibitem{bib:hama} Hamamatsu Photonics, Japan:
http://www.hamamatsu.com

\end{thebibliography}
\end{document}